\documentclass[aps,twocolumn,preprintnumbers,groupedaddress,nofootinbib,pra]{revtex4-2}
\pdfoutput=1

\usepackage[usenames,dvipsnames,table]{xcolor}
\usepackage{graphicx,amsmath,amssymb,amsthm,multirow,array,bm,bbm,esint}
\usepackage{enumitem}
\usepackage[mathscr]{eucal}
\usepackage[bbgreekl]{mathbbol}
\usepackage{epsf,amsfonts}
\usepackage{physics}
\usepackage{dsfont, mathtools}
\usepackage{slashed}
\usepackage{hyperref}
\usepackage{mathdots}
\usepackage{upgreek}
\usepackage{comment}

\usepackage[normalem]{ulem}

\hypersetup{
    pdfstartview={FitH},    %
    pdftitle={TITLE},    %
    pdfauthor={},     %
    colorlinks=true,       %
    linkcolor=blue,          %
    citecolor=red,        %
    filecolor=magenta,      %
    urlcolor=blue           %
}

\def\O{{\cal O}}

\def\eqr{\eqref}

\newcommand{\es}[2] {\begin{equation} \label{#1} \begin{split} #2 \end{split} \end{equation}}

\newcommand\mR{\mathbb{R}}
\newcommand\mZ{\mathbb{Z}}

\newcommand {\be} {\begin{align}}
\newcommand {\ee} {\end{align}}
\newcommand {\bes} {\begin {equation*}}
\newcommand {\ees} {\end {equation*}}
\newcommand {\beq} {\begin {equation}}
\newcommand {\eeq} {\end {equation}}
\newcommand {\bea} {\begin {eqnarray}}
\newcommand {\ea} {\end {eqnarray}}
\newcommand {\eea} {\end {eqnarray}}

\newcommand{\wormhole}[2]{%
  \pgfmathanglebetweenpoints{\pgfpointanchor{#1}{center}}{\pgfpointanchor{#2}{center}}%
  \let\whangle\pgfmathresult
  \draw[thick, blue!60!black]
    ($(#1.\whangle)+({\whangle+90}:0.35)$)
    .. controls ($(#1)!0.4!(#2)+({\whangle+90}:0.06)$) and ($(#1)!0.6!(#2)+({\whangle+90}:0.06)$) ..
    ($(#2.{\whangle+180})+({\whangle+90}:0.35)$);
  \draw[thick, blue!60!black]
    ($(#1.\whangle)+({\whangle-90}:0.35)$)
    .. controls ($(#1)!0.4!(#2)+({\whangle-90}:0.06)$) and ($(#1)!0.6!(#2)+({\whangle-90}:0.06)$) ..
    ($(#2.{\whangle+180})+({\whangle-90}:0.35)$);
}
\renewcommand{\es}[2] {\begin{equation} \label{#1} \begin{split} #2 \end{split} \end{equation}}

\newcommand{\vp}{\varphi}

\def\<{\langle}
\def\>{\rangle}

\def\ac{a_\text{crit}}

 \def\ie{\begin{equation}\begin{aligned}}
\def\fe{\end{aligned}\end{equation}}

\usepackage{upgreek}

\def\nn{\nonumber}

\def\Re{\text{Re}}
\def\Im{\text{Im}}

\def\cD{\mathcal{D}}

\def\Z{\mathbb{Z}}

\def\1{{\rm 1-loop}}

\def\Tr{{\rm Tr}}

\def\c{\cite}

\def\cM{\mathcal{M}}

\def\c{\cite}

\def\eqr{\eqref}
\def\O{{\cal O}}

\def\({\left(}
\def\){\right)}
\def\[{\left[}
\def\]{\right]}

\def\<{\langle}
\def\>{\rangle}

\def\vp{\varphi}

\def\cJ{\mathcal J}

\usepackage{tikz}

\begin{document}

\title{Positivity of the gravitational path integral implies the axionic weak gravity conjecture}

\author{Gabriele Di Ubaldo}
\affiliation{Leinweber Institute for Theoretical Physics and Department of Physics, University of California, Berkeley, CA 94720, USA,}
\affiliation{RIKEN iTHEMS Center for Interdisciplinary Theoretical and Mathematical Sciences,\\ 2-1 Hirosawa, Wako, Saitama 351-0198 Japan,}

\author{Luca V.\ Iliesiu}
\affiliation{Leinweber Institute for Theoretical Physics and Department of Physics, University of California, Berkeley, CA 94720, USA,}

\author{Henry W.\ Lin}
\affiliation{Joseph Henry Laboratories, Princeton University, Princeton, NJ 08544, USA,}

\author{Cynthia Yan}
\affiliation{Department of Physics, Harvard University, Cambridge, MA 02138, USA.}

\begin{abstract}
The gravitational path integral can compute inner products between different states of open and closed universes. To have a well-defined Hilbert space, these inner products should be positive semi-definite, which is not manifest in the low-energy effective theory. In this letter, we analyze the constraints that the positivity of inner products imposes on gravitational theories coupled to axions. If the axion has an exact shift symmetry, we show that, under mild assumptions, a combined positivity constraint on closed and open universes is violated when one includes certain wormholes. 
In low-energy effective theories where these wormholes are perturbatively stable, positivity requires that the wormholes have a non-perturbative instability that breaks the shift symmetry.  This leads to a sharp version of the axion weak gravity conjecture, including precise numerical constants. We relate the bound to possible extensions of other swampland conjectures, arguing for an imaginary continuation of the distance conjecture. We comment on how the bound applies to axions in string theory and discuss phenomenological implications. 
\end{abstract}

\maketitle

\section{Introduction}
\label{sec:intro}

Which low-energy effective field theories (EFTs) of gravity can be embedded in self-consistent theories of quantum gravity? %
In this letter, we constrain the properties of  low-energy theories of gravity in flat space and AdS${}_{d+1}$ with axions  by imposing that all linear combinations of states prepared by the path integral have positive norm.

An axion is a periodic scalar with a continuous shift symmetry, which is believed to be broken by non-perturbative effects. However, the scale at which these effects become important -- and at which the shift symmetry is effectively broken -- is, in general, unknown. Because any consistent theory of quantum gravity is believed not to have exact global symmetries \c{Banks:2010zn, Misner:1957mt, Abbott:1989jw, Kallosh:1995hi,Polchinski:2003bq, Harlow:2018tng, Harlow:2018jwu, Heidenreich:2020pkc, Chen:2020ojn, Yonekura:2020ino, Harlow:2020bee, Hsin:2020mfa, Belin:2020jxr}, an axion coupled to gravity should not have an exact shift symmetry, and the scale at which non-perturbative effects break the symmetry is expected to lie below the Planck scale. A sharpening of this expectation is the axionic weak gravity conjecture (WGC) \c{Arkani-Hamed:2006emk,Rudelius:2015xta,Brown:2015iha,Heidenreich:2015wga,Reece:2023NoGlobalSymmetries,Arvanitaki:2010StringAxiverse,Svrcek:2006AxionsStringTheory,Cicoli:2012IIBAxiverse}, which states that, for any axion with decay constant $f_a$, there must exist an instanton with action
\begin{equation}
\label{eq:WGCbound}
S_{\text{inst}} \lesssim \O(1)\, \frac{M_{\text{Pl}}}{f_a}\,.
\end{equation}
The goal of this letter is to give an argument for \eqref{eq:WGCbound}, including precise $\O(1)$ factors, from the positivity of the gravitational path integral. 

We start by showing that a low-energy theory of gravity with an axion with an exact axion shift symmetry fails to define a good Hilbert space. The gravitational path integral can be used to compute inner products between different states of open and closed universes \c{Coleman:1988cy,Marolf:2020xie}; for these inner products to give rise to a well-defined Hilbert space, they must be positive semi-definite -- a property that is not manifest in the low-energy EFT. Axion wormholes \cite{Giddings:1987cg,Giddings:1988cx,  Lee:1988ge, Abbott:1989jw, Coleman:1989zu, Coleman:1988cy, Tamvakis:1989aw, Burgess:1989da, Kallosh:1995hi, Bergshoeff:2004pg,ArkaniHamed:2007js,Bergman:2009AdSWormholes,Gupta:1990MagneticWormholes,Hertog:2017owm,Andriolo:2022MassiveDilaton,Jonas:2023ZooAxionicWormholes,Marolf:2025StabilityRevisited,Alvey:2021AxionQuality, Loges:2023ypl, Witten:2026twr,Held:2026huj,Held:2026JTImaginaryScalars} are one of the geometries that contribute to such inner products. However, the amplitude of such wormholes becomes unbounded when we complexify the axion field and tune the difference between the imaginary boundary values of the axion on the wormhole mouths past the following values:
\bea
\label{eq:distance-conjecture-imaginary-bound}
&\textrm{AdS:} \quad |\Im \,\Delta a| \leq  \frac{1}{f_a} \sqrt{\frac{\pi (d-1)}{8d\,G_N}} \equiv \Im \,\Delta \ac^\text{AdS}\,,\\ 
&\textrm{Flat space:} \quad |\Im \,\Delta a| \leq  \frac{1}{f_a} \sqrt{\frac{\pi d}{8(d-1)\,G_N}} \equiv \Im \,\Delta \ac^\text{flat}\,.\nn
\eea
This bounds the maximal field excursion that the axion can take in the imaginary direction before wormholes can proliferate in any given gravity solution. Due to its similarity with the distance conjecture, we call \eqref{eq:distance-conjecture-imaginary-bound} the Imaginary Distance Conjecture. The bound is universal: it does not depend on the AdS scale or the curvature of the spatial slices of the wormhole. The divergence that we encounter when saturating \eqref{eq:distance-conjecture-imaginary-bound} is even more problematic than causing wormholes to proliferate: it leads to a violation of the combined positivity constraint on closed and open universes, when %
such wormholes are perturbatively stable (see \cite{Hertog:2024nys,Marolf:2025StabilityRevisited}) and contribute to the path integral. 

 To cure the negativity of norms that we encounter as we saturate \eqref{eq:distance-conjecture-imaginary-bound}, we argue that the wormhole should be non-perturbatively unstable under the production of instantons that break the shift symmetry for boundary values of the axion satisfying \eqref{eq:distance-conjecture-imaginary-bound}.\footnote{In fact, in such a case, the one-boundary saddle will also become non-perturbatively unstable under the proliferation of instantons.}  Quantifying the strength of such corrections by the single-instanton action $S_{\text{inst}}$, we find that positivity only has a chance of being restored if\footnote{\label{footnote:AdS-vs-flatspace}Our arguments are most clear in AdS since the argument uses the existence of wormholes with $S^1 \times \cM$ slicing. Such solutions do not exist in flat space, where we can only find saddles for sphere slicing. Nevertheless, imposing that the wormhole becomes unstable before the divergence occurs yields a similar bound to \eqref{eq:boundintro} in flat space.}
\bea
\label{eq:boundintro}
   S_{\text{inst}} \leq  \frac{1}{2f_a} \sqrt{\frac{\pi (d-1)}{8d\,G_N}}
\eea
When the inequality \eqr{eq:boundintro} is saturated, instantons become strong enough to trigger a non-perturbative instability of the wormhole, and their proliferation can drastically modify the one-boundary and two-boundary amplitudes to restore positivity. The resulting bound \eqr{eq:boundintro} takes precisely the form of the axionic WGC \eqr{eq:WGCbound}, which we are thus able to derive -- including its precise numerical constants (see also \cite{Hebecker:2018ofv_review, Brown:2015iha,   Montero:2015ofa, Heidenreich:2015wga, Hebecker:2017uix, Hebecker:2016dsw, Heidenreich:2021xpr,Andriolo:2022rxc,  Heidenreich:2024dmr}) -- from the sole requirement that the theory admits a consistent Hilbert space. In a phenomenological setting, where the axion couples to the topological term of the Yang-Mills field, $\Tr F\wedge F$, and one can use an estimate for the instanton partition function in QCD to obtain an upper bound on the axion decay constant $f_a \lesssim O(10^{17})$ GeV, see equations \eqref{eq:fa-estimate} and \eqref{eq:fa-qcd-bound2}.

 The remainder of this paper is organized as follows. In section~\ref{sec:positivity-rules}, we summarize the general positivity constraints on single- and 
multi-boundary contributions to the gravity path integral. In 
section~\ref{sec:wormhole-amplitudes}, we compute wormhole amplitudes in AdS$_{d+1}$, discuss their holographic interpretation, and 
derive their universal divergence when the bound~\eqref{eq:distance-conjecture-imaginary-bound} is 
saturated. Building on this, in section~\ref{sec:arg-for-weak-gravity-bound}, we present our 
argument for the axion weak gravity bound in \eqref{eq:boundintro}. Finally, in 
section~\ref{sec:discussion}, we analyze the resulting phenomenological bound on 
the axion decay constant, how our reasoning applies to axion wormholes in string theory, and possible generalizations of our results for other swampland conjectures. 

\noindent\textbf{Note added.} As this work was being completed, we became aware of the concurrent work \cite{Maldacena:2026ImaginaryDistanceBound}, which has partial overlap with the results presented here.

\section{Positivity rules}
To start, we discuss the minimal positivity requirements that any gravitational path
integral must satisfy if it prepares states in a genuine Hilbert space. There are two types of states the GPI can prepare: open universe and closed universe states. The
path integral with one open asymptotic boundary and boundary condition $\cJ$ prepares an
``open-universe'' state $\ket{\cJ}$.  Gluing two such boundaries computes the
inner product
\bea
\label{eq:inner-prod-open}
    \langle \cJ_1|\cJ_2\rangle
    =
    Z(\cJ_1^\Theta \cup \cJ_2)\,,
\eea
where $\Theta$ denotes the reflection, or CPT conjugation, appropriate to the
chosen boundary condition, $\cJ_1^\Theta \cup \cJ_2$ denotes the boundary condition on a single closed boundary obtained by gluing the two open boundaries, and where $Z(J)$ denotes the GPI sum over geometries satisfying the boundary condition $J$. Similarly, the state $\ket{J}$ in the closed universe Hilbert space is obtained by preparing closed asymptotic boundaries with boundary condition $J$. The inner product is computed by the path integral 
\bea 
\label{eq:inner-prod-closed}
\braket{J_1}{J_2} = Z(J_1^\Theta \cup J_2)\,,
\eea 
where $J_1^\Theta \cup J_2$ are the boundary conditions obtained through the union of the closed universe boundaries specified in the bra and ket.  
Positivity of the gravitational Hilbert space implies that any Gram matrix involving closed or open universe inner-products must be positive semi-definite \cite{Marolf:2020xie, IliesiuTalks, Liu:2025ikq, toAppearPositivity},  
\bea 
\label{eq:positivity-condition}
\braket{\cJ_1}{\cJ_2} \succeq 0\,, \qquad\braket{J_1}{J_2} \succeq 0  \,.
\eea 
This condition has a simple interpretation when considering asymptotic boundaries, which, after gluing in the case of open universes, consist of a set of connected components with topology $S^1\times \cM$ with homogeneous boundary conditions along $S^1$ which we will generically denote by $J$.  In such
cases, \eqref{eq:inner-prod-open} and \eqref{eq:inner-prod-closed} can be written as a Laplace transform, for example
 \bea
 \label{eq:inner-prod-open-and-closed-example}
     &\braket{\cJ_1}{\cJ_2} =  \int \mathrm{d}E \,\,\overline{\rho_J(E)} e^{-(\beta_1 E^* + \beta_2 E) },\\
       &\braket{J_1}{J_2} = \int \mathrm{d}E_1\, \mathrm{d}E_2 \,\overline{\rho_J(E_1)\rho_J(E_2)}\, e^{-(\beta_1 E^*_1 + \beta_2 E_2) },\nn
    \eea
when $\cJ_i$ and $J_i$ only contain a single connected component.\footnote{A similar rewriting can be done when $\cJ_i$ (or $J_i$) have $n$ (or $n/2$) connected components in terms of the inverse Laplace-transform $\overline{\rho(E_1)\dots \rho(E_n)}$.} Above, $E_1$ and $E_2$ are the eigen-energies of the ADM Hamiltonian. Since we have not assumed that this operator is Hermitian, the energies can be complex, and the integration range in such cases is over the entire complex plane. The positivity condition \eqref{eq:positivity-condition} can be shown (see Appendix \ref{positivity}) to then imply that 
  \bea
\overline{\rho_J(E_1)\cdots\rho_J(E_n)} = \int \mathrm{d}\alpha\, P(\alpha) \rho^\alpha_J(E_1)\cdots\rho^\alpha_J(E_n),
    \eea
with
    \bea
    \label{eq:positivity-measure-and-DOS}
    P(\alpha)\geq 0, \qquad \rho^\alpha_J(E) \geq 0\,.
    \eea
Positivity thus allows us to interpret $\rho_\alpha(E) $ as densities of states in an ensemble (or in whatever
coarse-grained description the GPI represents) labeled by $\alpha$, $P(\alpha)$ as a probability measure, and $\rho_\alpha(E_1)\dots\rho_\alpha(E_n)$ as computing the $n$-th moment in this ensemble. When $\rho_\alpha(E)$ only has support on the real line, then inner products such as \eqref{eq:inner-prod-open-and-closed-example} capture moments of the partition function with sources turned on, i.e.,  $\braket{\cJ_1}{\cJ_2} = \overline{Z(J, \beta_1+\beta_2)} $ and $\braket{J_1}{J_2} = \overline{Z(J, \beta_1)Z(J, \beta_2)} $. Since $\rho_\alpha(E)\geq 0$, all partition functions are also positive random variables in the ensemble whose moments satisfy the Stieltjes moment problem. Denoting $m_n(J, \beta) = \overline{Z(J, \beta)^n}$, positivity implies

\bea
    \left(m_{i+j}\right)_{i,j=0}^{N}\succeq0,
    \qquad N=0,1,\ldots,
    \nonumber\\
    \left(m_{i+j+1}\right)_{i,j=0}^{N}\succeq0,
    \qquad N=0,1,\ldots .
    \label{eq:stieltjes-hankel}
\eea
The first matrix is the usual Hamburger positivity condition.  The second,
``shifted'' Hankel matrix is the extra requirement that the measure has support
only on $Z_\alpha(J, \beta)\geq0$. It is instructive to write the lowest inequalities in terms of the cumulants of the distribution, $\kappa_n = \overline{Z(J, \beta)^n}_c$, where $\overline{(\dots)}_c$ captures the contribution of fully connected geometries to the GPI:
\bea
    \kappa_1\geq0,\qquad
    \kappa_2\geq 0,\qquad
    \kappa_1 \kappa_3\geq \kappa_2(\kappa_2 - \kappa_1^2),
    \label{eq:stieltjes-basic}
\eea
with further inequalities involving all higher cumulants. The last inequality proves particularly useful: it restricts the two-boundary wormhole amplitudes ($\kappa_2$), which appear on the RHS of the inequality, from becoming unbounded. As we advertised in the introduction and will justify shortly, the axion wormhole amplitude with Dirichlet boundary conditions becomes divergent at the critical value in \eqref{eq:distance-conjecture-imaginary-bound} which violates the inequality and leads to a violation of positivity.

\label{sec:positivity-rules}
\section{Wormhole amplitudes} 
\label{sec:wormhole-amplitudes}

We now compute the semiclassical wormhole amplitudes that enter the inequality in \eqref{eq:stieltjes-basic}.
We consider an axion $a\sim a+2\pi$ coupled to gravity in
$D=d+1$ Euclidean dimensions with $d\ge 2$\footnote{In this letter, we will not consider the $d=1$ JT case \cite{Maldacena:2018lmt, Garcia-Garcia:2020ttf}, which has no divergence in the wormhole amplitude at any finite value of the couplings, in contrast to the cases $d \ge 2$. We will discuss this further in \cite{DiUbaldo:AxionWormholesLong}.}, 
\bea
        \label{eq:action}
 S_{EFT}= \int \mathrm{d}^{D}x \sqrt{g}\left(-\frac{R-\Lambda%
 }{16\pi G_N}+\frac{f_a^2}{2}\partial_\mu a \partial^{\mu}a\right).
\eea
Such an EFT does not generically have a multi-boundary saddle point with Dirichlet boundary conditions for the axion when fixing the axion to be real at the boundary.\footnote{There exists an axion wormhole saddle for real axions in AdS with hyperbolic slicing. The analysis that we perform below also applies for this case and therefore we will not treat it independently.} Because of that, it is often convenient to compute the wormhole at fixed charge under the shift symmetry, where saddle-point configurations do exist, and only afterwards transform
to a fixed boundary value of the axion by summing over all possible fluxes. To compute the amplitude of such wormholes, we can dualize the axion to a $(d-1)$-form field: fixing the charge under the shift symmetry is equivalent to fixing the flux of the higher form field, requiring Dirichlet boundary conditions in this dual theory. For real fluxes, we find that such solutions are entirely real; if such wormholes are perturbatively stable, they should therefore contribute to the GPI. In  the scalar description, finding the flux amounts to
working with the complex saddle $a\in i \mR$, for which the axion kinetic term has
the sign needed to support a Euclidean wormhole.  %
Denoting the corresponding integer flux by $m$, the
fixed-flux wormhole action grows linearly at large $|m|$ (see Appendix \ref{solution}),
\bea
    S_{\rm WH}(m)
    = |m| (\Im \Delta a_\text{crit}) +
 \cdots\,,
    \label{eq:flat-fixed-flux-action}
    \eea
with $\Im \Delta a_\text{crit}$ defined in \eqref{eq:distance-conjecture-imaginary-bound} for both flat space and AdS. 
The $\dots$ denote terms that are
subleading at large flux, including the one-loop determinant and curvature-
dependent corrections.  The important point is that the coefficient of $|m|$ is
universal: it does not depend on the AdS radius and, at leading order, it is independent of the slicing of the wormhole.\footnote{One can consider spatial manifolds besides $S^1\times \cM$. } At large $m$, the saddle-point description is highly robust: all curvature invariants are suppressed in $m$, making the EFT in \eqref{eq:action} highly trustworthy. %

To impose Dirichlet boundary conditions for the physical axion, one sums over
flux sectors with the Fourier weight $e^{i m \Delta a}$, where
$\Delta a=a_1-a_2$ is the difference between the two boundary values \cite{Witten:2026twr}. For example, for flat space wormholes with spherical slicing and AdS wormholes with flat slicing, this gives\footnote{As in \cite{Chen:2020ojn,Hsin:2020mfa}, the fact that the wormhole has a non-trivial dependence on $\Delta a$ suggests that the bulk global shift symmetry is only a symmetry of the ensemble but not a global symmetry for each individual member of the ensemble.  }
\bea
  Z_\text{WH}(\Delta a)
  &=
  \sum_{m \in \mathbb Z} Z^\text{1-loop}_m
  \exp\left[- {|m|} (\Im \Delta \ac)+i m \Delta a\right] \nonumber \\
  &\sim
  \frac{\sinh\left(\Im \Delta \ac\right)}
  {\cosh\left(\Im \Delta \ac\right)-\cos(\Delta a)}\,.
  \label{eq:flat-wormhole-flux-sum}
\eea
where, for concreteness, in the last equality, we have neglected the one-loop determinant. 
The flux sum makes sense even when there is
no real Dirichlet saddle; in this sense, it provides an off-shell derivation of the Dirichlet wormhole amplitude by starting with the on-shell fixed flux wormhole.

This expression also displays the aforementioned divergence regardless of the exact form of the one-loop determinant.  For imaginary axion separation, the Fourier weight becomes exponential rather than
oscillatory.  The large-flux tail then converges only while
\bea
    |\Im \Delta a|
    < \Im \Delta \ac\,.
    \eea
At the boundary of this strip, the connected two-boundary amplitude becomes
singular.\footnote{Physically, this divergence can be attributed to the fact that the two wormhole boundaries come close to one another. For instance, in AdS, the renormalized proper length between the two boundaries can become arbitrarily negative at large flux.}

In holographic language, an AdS axion is dual to an exactly marginal
periodic coupling of the boundary theory. For instance, in the context of $\mathcal N=4$ SYM fixing the axion at the boundary fixes the $\theta$ angle; alternatively, setting the flux of the axion fixes the instanton number of the boundary theory.  The wormhole, therefore, computes a
connected contribution to the statistical correlation of partition functions at 
two values of this coupling. To be precise, if the partition function is invariant under $\theta \to -\theta$ due to the theory with $\theta = 0$ being time-reversal invariant, the second moment of the partition function is given by a sum of two wormhole contributions\footnote{This is because the wormholes in the $\theta =0$ theory can include a reversal of orientation which changes the sign of the flux one of the sides of the wormhole. }
\bea 
\overline{Z(\theta_1) Z(\theta_2)}_c =  Z_\text{WH}(\theta_1-\theta_2) + Z_\text{WH}(\theta_1 + \theta_2)\,.
\eea 
The divergence encountered at the critical imaginary value $\Im \Delta\ac$ says that the statistical
correlation between the two boundary quantities diverges. While this might be reasonable if the partition function itself diverges, if it does not, such a divergence is not physically reasonable and leads to a violation of positivity that we outline below.\footnote{A probability distribution with a divergent standard deviation is mathematically possible. Positivity constrains higher cumulants of the distribution, coming from wormholes with $n\geq3$ boundaries, to have even stronger divergences than the two-boundary wormhole contribution. To rescue positivity, we need a divergence at least as strong as $\kappa_{2n+1} \sim (\kappa_n)^{2}$.}

 \section{An argument for the weak gravity bound}
\label{sec:arg-for-weak-gravity-bound}

We now explain how the divergence above leads to the weak gravity bound.  Let us start with the low-energy theory \eqref{eq:action} that has an exact shift symmetry for the axion field, $a \to a+c$.
In the ensemble interpretation mentioned above, this shift symmetry is a symmetry of
the ensemble, with the axion boundary value fixing the boundary $\theta$ angle, i.e., $\overline{Z(\theta)} = \overline{Z(\theta+c)}$.

For complex $\theta$, the boundary theory is not manifestly
unitary; however, the theory with $\theta=0$ is unitary. Due to the bulk shift symmetry we therefore have $\overline{Z(\theta)} = \overline{Z(0)}$ and $\overline{\rho_\theta(E)} = \overline{\rho_{\theta = 0} (E)}$ with support only for real $E$. Eq.~\eqref{eq:positivity-measure-and-DOS}, therefore, 
implies that, even for complex $\theta$, each member of the ensemble has a density of states with support solely for real $E$. The associated partition function is a non-negative
random variable ($Z_\alpha(\theta) \geq 0$) whose moments must therefore obey the Stieltjes
constraints \eqref{eq:stieltjes-hankel}.

Near the critical imaginary axion separation, the two-boundary wormhole predicts
a very large connected second moment.  If wormholes connecting more than two boundaries are
suppressed, the statistics of
$Z_\alpha(\theta)$ are those of a
real Gaussian random variable once $\theta$ is purely imaginary.\footnote{The suppression of higher connected moments is a good approximation for the spectral form factor, which is approximately Gaussian as it is the sum of many random phases \cite{Saad:2018bqo}.}  As
$|\Im\,\Delta a|$ approaches the critical value, the variance grows without bound
while the one-boundary mean remains finite since $\overline{Z(\theta)} = \overline{Z(\theta =0)}$ and  $\overline{Z(\theta =0)}$ is finite.  A Gaussian with standard deviation much
larger than its mean is negative with probability approaching $50\%$,
which contradicts the conclusion that $Z_\alpha\geq0$ 
with probability 1.

To improve the above argument, we can consider corrections from multi-boundary wormholes whose contribution we expect to be suppressed by $e^{-\frac{\#}{G_N}}$ compared to the two-boundary wormhole. To check positivity, we can analyze the last Stieltjes constraint in \eqref{eq:stieltjes-basic} as $\Im\,\theta \to |\Im \Delta a|/2$: in this regime, $\kappa_1 = \overline{Z(\theta )}=  \overline{Z(\theta =0)}$ remains finite, $\kappa_2 = \overline{Z(\theta )^2}_c$ is arbitrarily large, and $\kappa_3 = \overline{Z(\theta)^3}_c$ is suppressed by $e^{-\frac{\#}{G_N}}$ relative to the one- and two-boundary contributions. Therefore, small corrections due to multi-boundary wormholes cannot restore positivity.\footnote{In 3d gravity, the contribution of multi-boundary wormholes with higher genus boundaries and Dirichlet boundary condition for a free scalar was conjectured in \cite{chandra2024e}, by assuming a Gaussian ensemble for the OPE coefficients of a putative boundary ensemble. When postulating such an ensemble interpretation, the Stieltjes positivity condition is not automatically satisfied, and we have explicitly checked that it is violated. }  When $\kappa_2$ is parametrically large, the Stieltjes moment problem can only be satisfied if higher connected wormholes grow in a correlated and
equally singular way, a feature that we do not expect in the $G_N^{-1}$ expansion.\footnote{For instance, we would need $\kappa_3 \sim \Omega(\kappa_2^2)$ as we approach the divergence. While we expect the three-boundary wormhole to have a similar divergence to the two-boundary wormhole once two of the boundaries come close to one another (giving $\kappa_3 \sim e^{-\#/G_N}\kappa_2 $ as we approach the divergence) we do not expect a faster divergence to be possible.  }  The natural conclusion is that the saddle responsible for the divergent second moment is not a valid
contribution in a theory of gravity with a well-defined Hilbert space of open and closed universes.\footref{footnote:AdS-vs-flatspace}

There are two logical possibilities.  One could declare that all axion
wormholes should be excluded from the gravitational path integral by a rule not
visible in the low-energy theory.\footnote{See \cite{Witten:2026twr,Held:2026huj} for this perspective.} Given that the wormhole amplitude is obtained by summing over real saddles that are expected to be perturbatively stable this is difficult to imagine.\footnote{One would also have to exclude axion wormholes with hyperbolic slices that are smoothly connected to the Maldacena-Maoz wormholes whose statistical meaning was discussed in $3d$ gravity.}    The more conservative interpretation is that
while these saddles are perturbatively stable, they must exhibit a non-perturbative
instability. The instability must come from effects that break the shift symmetry.  We will
call these effects instantons, with the understanding that in a UV completion, they may be realized by the appropriate Euclidean branes or gauge instantons. %
If all such instantons remain suppressed as $\Im \,\theta \to  |\Im \Delta \ac|/2$, then the Stieltjes positivity of the density of states is still violated. Therefore, to have a chance to restore positivity, instantons must be able to proliferate by having a negative contribution to the effective action for $|\Im \,\theta| <  |\Im \Delta \ac|/2$.  The real contribution of a single instanton of action
$S_{\rm inst}$ in a background with complex axion value $a$ to the real part of the effective action is
\bea
   \Re\, S_{\rm eff}=S_{\rm inst}\mp \Im\, a\,,
\eea
where the sign distinguishes instantons from anti-instantons.  For instantons to proliferate in at least some region of the wormhole background, we therefore need
\begin{align}
    S_{\rm inst}
    &\le
    \frac{1}{2}\,|\Im \,\Delta \ac|\label{eq:wgc-bound-from-positivity}\\
    &= \begin{cases}
        \textrm{ Flat space:}\,\,\frac{1}{2f_a}\sqrt{\frac{\pi d}{8(d-1)\,G_N}}, \nn \\    \textrm{ AdS:}\,\,
        \frac{1}{2f_a}\sqrt{\frac{\pi(d-1)}{8d\,G_N}}\,.
    \end{cases}
\end{align}
which reproduces the bound \eqref{eq:boundintro}. Note that when \eqref{eq:wgc-bound-from-positivity} is satisfied, instantons will also proliferate on the one-boundary saddles for $|\Im \,\theta| <  |\Im \Delta \ac|/2$. This makes it possible for the problematic Stieltjes constraint in \eqref{eq:stieltjes-basic} to be satisfied since both $\kappa_1$ and $\kappa_2$ receive large corrections once the instantons proliferate.

Above, we are using the term ``instanton'' loosely. If the background is AdS $\times \cM$, with $\cM$ a (possibly large) compact dimensions, the ``instanton'' just needs to be somewhat localized in the AdS dimensions; for example, it could be a brane wrapping some of the compact dimensions.

\section{Discussion}
\label{sec:discussion}

We have argued that the gravitational path integral around a low-energy theory of gravity coupled to an axion with an exact shift symmetry cannot define a Hilbert space with positive-norm states. Demanding that the wormholes that lead to the violation of positivity have an instanton-induced instability reproduces the axionic weak gravity bound \eqref{eq:WGCbound}, including its precise $\O(1)$ coefficient. 

We emphasize that our analysis identifies \emph{when} positivity fails under the assumption that multi-boundary wormholes are subleading in the $G_N^{-1}$ expansion, a natural assumption that is however, difficult to check in higher dimensions without explicitly finding such saddles. Furthermore, we have not proved that positivity is in fact restored once instantons proliferate, but this is not necessary to derive the axionic weak gravity bound. Showing that the proliferation of instantons explicitly modifies the one- and two-boundary amplitudes in such a way that all Stieltjes constraints \eqref{eq:stieltjes-hankel} are satisfied is an interesting open problem. It is possible that further modifications of the EFT are required for the gravitational path integral to give rise to a well-defined Hilbert space. 

More broadly, our result illustrates that positivity of the gravitational path integral is a sharp diagnostic for the consistency of low-energy EFTs of gravity. Most existing swampland conjectures are motivated by indirect arguments or by patterns observed in string compactifications \cite{Vafa:2005ui,Agmon:2022thq}; here, instead, the axionic WGC arises from a single requirement -- the existence of a positive-norm Hilbert space -- applied to an explicit semiclassical computation. We hope that this is the first step in a broader program of bootstrapping low-energy EFTs of gravity by demanding positivity of all path-integral inner products.\\

\textbf{Axions in string theory. } Many axions in string theory couple non-trivially to a dilaton through derivative interactions that preserve the shift symmetry, giving an action of the form \cite{ArkaniHamed:2007js}:
\bea 
S_\text{EFT}
 \supset \frac{1}{32\pi G_N} 
\int \! \mathrm{d}^{D} \! x \sqrt{g}%
\left[(\partial_\mu \varphi)^2- e^{-\beta\vp}(\partial_\mu a)^2\right]\,.
\label{eq:axio-dilaton-action}
\eea
For such axions, the wormhole analysis of Sec.~\ref{sec:wormhole-amplitudes} carries over: the axio-dilaton wormhole amplitude diverges when the imaginary part of the axion exceeds a critical value, signaling a violation of positivity in low-energy theories of supergravity (such as type IIB  supergravity) that must be cured non-perturbatively.

In type IIB string theory, the relevant non-perturbative effects are D$(-1)$-brane instantons and anti-instantons. %
Demanding that this D$(-1)$-instability sets in before the axio-dilaton wormhole amplitude diverges yields a sharp lower bound on the axion-dilaton coupling, $|\beta|>\beta_{\rm crit}=\sqrt{2d/(d-1)}$. For AdS$_5\times S^5$ this requires $\beta_{\rm crit}=\sqrt{8/3}$, comfortably below the type IIB value $\beta_{IIB}=-2$. We perform an analysis for such theories in App.~\ref{axiodilaton}.\\

\textbf{Connections to other swampland conjectures.} \textit{The distance conjecture. } The bound \eqref{eq:distance-conjecture-imaginary-bound}, which we have argued is required for positivity of the gravitational path integral, can be viewed as a new distance bound for scalar field excursions in the \emph{imaginary} direction. This is a ``time-like'' direction in moduli space. It is qualitatively distinct from the standard distance conjecture \c{Ooguri:2006in}, which states that an infinite tower of light states emerges as one moves an infinite proper distance in moduli space along \emph{real} field directions. Our bound is not a statement about an emerging tower but rather about a divergence in the gravitational path integral at a \emph{finite} imaginary distance set by the Planck scale and the axion decay constant. In the real direction, by contrast, axion wormholes do not diverge, and our argument has nothing direct to say about the standard distance conjecture.

That said, the axionic weak gravity bound \eqref{eq:wgc-bound-from-positivity} we obtained by imposing positivity does %
constrain the behavior of the field along real directions. The instanton action %
controls the maximum periodicity that the axion can have in the real direction; if $S_\text{1-inst} > 1$ then the period of the axion is bounded in the real direction by the Planck scale. 

It would be interesting to extend our analysis to non-compact scalars. The wormhole computation of Sec.~\ref{sec:wormhole-amplitudes} applies %
and we encounter the same divergence at imaginary field separation when integrating over the possible fluxes, which are now continuous. This raises two questions: which potential deformations are compatible with positivity of the gravitational path integral, and does the resulting bound translate into a constraint on real-direction field excursions analogous to the standard distance conjecture?

\textit{The weak gravity conjecture.} Axions can be obtained by dimensionally reducing a higher-dimensional $U(1)$ gauge field on a circle and are given by the holonomy around the circle. In flat space, since the radius of the circle is determined dynamically, the resulting theory after dimensional reduction takes the form \eqref{eq:axio-dilaton-action} for which we can explicitly find axio-dilaton wormholes. %
Instantons arise from charged particles winding around the circle, and the positivity bound from the axio-dilaton wormhole becomes the ordinary weak gravity conjecture \cite{Arkani-Hamed:2006emk, Harlow:2022gzl}, including the precise $\O(1)$ coefficient:
\bea 
\frac{m}{|q|}\leq  \sqrt{\frac{e_{d+2}^2d}{8\pi G_{d+2}(d-1)}} . %
\eea 
In the above $e_{d+2}$ is the Maxwell coupling constant in the higher dimensional theory.
Therefore, our derivation of the axionic WGC seemingly applies to derive the ordinary WG bound through a dimensional uplift.

There is, however, a subtlety: in such setups, our assumption that the one-boundary answer is approximately shift symmetric fails even if the axionic theory arising from dimensional reduction is shift invariant. This happens when including saddles in the GPI in which the circle on which the dimensional reduction is performed becomes contractible, to form a Euclidean cigar. The holonomy at the tip of the cigar is constrained, $\oint A \in 2\pi \mZ$. Since the holonomy at the tip needs to have this special value in order for the gauge configuration to be smooth, the shift symmetry of the dimensionally reduced axion is broken in such configurations.\footnote{This happens for the one-boundary AdS instantons \cite{Witten:1998uka,Barbon:1999zp,Bigazzi:2015bna} whose action explicitly depends on the $\theta$-angle and approximates the $\theta$ dependence of the large $N$ Yang-Mills vacuum \cite{Aitken:2018YMTheta,Vicari_2009}.}  This issue is, however, straightforward to overcome: if the one-boundary density of states still has support for real $E$ when $\theta$ is imaginary and if the one-boundary partition function is still convergent for $\Im \theta \leq |\Im \Delta \ac|/2$ then the Stieltjes constraints are still violated. Consequently, the presence of charged particles is required in order to have a chance to restore positivity.  \\

\textbf{A bound on $f_a$ in our universe.} 
Consider a $d=3$ bulk theory with an axion coupled to Yang-Mills theory,
\bea
\label{eq:Yang-Mills-action}
S_{\textrm{YM}} = - \int \! \frac{1}{2g_{\textrm{YM}}^{2}}\, \operatorname{tr}(F \wedge \star F)  + \frac{i a}{8\pi^{2}} \operatorname{tr}(F \wedge F)
\eea
with the axion normalized to have period $2\pi$.  Expanding around the sector
with vanishing gauge field strength, the leading non-perturbative corrections
come from Yang-Mills instantons with instanton number $\pm 1$.  For imaginary
axion $a=i a_W$, their effective action is
\bea
S_{\rm eff}^{\pm}
=
-\log Z_{1\text{-inst}}(g_{\textrm{YM}})\pm a_W\,,
\eea
where $Z_{1\text{-inst}}(g_{\textrm{YM}})$ is the one-instanton contribution evaluated
without the topological term.  Thus, one of the two signs becomes unsuppressed
when the imaginary axion is large enough.  Requiring this to happen before the
flat-space wormhole diverges gives
\bea
    -2\log Z_{1\text{-inst}}(g_{\textrm{YM}})
    <
    \frac{1}{4f_a}\sqrt{\frac{3\pi}{G_N}}\,.
    \label{eq:qcd-instability-bound}
\eea
Using the rough Standard Model estimate $S_{\rm 1-inst } \equiv \log Z_{1\text{-inst}}(g_{\textrm{YM}})\simeq 12$ (which was evaluated at the scale of the average instanton size $\rho \sim 1/3 \text{ fm}$ at low energies) for
the $SU(3)$ Yang-Mills instanton action, gives  a bound of $f_a$ (see \cite{Benabou:2025kgx} for a comparison)
\bea
f_a \lesssim 3.9 \times 10^{17}\,{\rm GeV}\,.
\label{eq:fa-estimate}
\eea
With some phenomenologically viable assumptions about the neutrino spectrum, the standard model (with a QCD axion) also has an AdS$_3 \times S_1$ compactification, where the $S_1$ has a radius $R_0$ that is of order 3 microns \cite{Arkani-Hamed:2007ryu}. Applying our AdS bound with $d=2$ gives
\bea
 f_a^{D=4}
<
\frac{1}{8S_{\text{1-inst}}}\sqrt{\frac{\pi}{G_N^{D=4}}}\, \sim 2.3 \times 10^{17} \,{\rm GeV}\,,
\label{eq:fa-qcd-bound2}
\eea
where we have used that $f_a^{D=3} = (2\pi R_0)^{1/2} f_a^{D=4}$ and $G_N^{D=3} =  G_N^{D=4}/(2\pi R_0)$.

  The precise number should not be taken as a sharp phenomenological bound, since it depends on the
definition of the one-instanton contribution (which only makes sense in the dilute gas approximation) and on the scale at which the gauge
coupling is evaluated.  The main point is that in the absence of other non-perturbative effects in \eqref{eq:Yang-Mills-action}, the positivity problem only has a possible resolution if ordinary Yang-Mills instantons become
important before the axion wormhole amplitude diverges.

Since the dilute gas approximation is inaccurate at low energies, it seems likely that one can improve on this estimate \eqref{eq:fa-estimate} by using the chiral perturbation theory estimate of the instanton potential. Given some instanton potential to any axion field, one could (at least in principle) recompute the moments of the partition function \eqref{eq:stieltjes-hankel} and test whether positivity is satisfied. It would be interesting to carry out this {\it wormhole bootstrap} in detail \cite{DiUbaldo:AxionWormholesLong}.
Alternatively, the bound can also become more reliable if one had a wormhole background in which the relevant instanton action is evaluated at a higher energy scale, where the negativity problem persists. If the wormhole carries non-trivial boundary conditions for additional Standard Model fields, the corresponding sources can contribute to the energy density localized near the regions of largest imaginary axion. If such sources do not modify the behavior of the wormhole amplitude at large flux, the negativity problem would persist. The relevant Yang-Mills instantons would then be small and weakly coupled, and one could evaluate $S_{YM,\rm inst}$ at a short-distance scale where the dilute-gas approximation is under control. It would be interesting to understand whether this scale can be pushed all the way up to the GUT scale, %
making the resulting bound on $f_a$ tighter by roughly an order of magnitude relative to \eqref{eq:fa-estimate}.\\

\begin{acknowledgments}
We thank Victor Ivo, Jorrit Kruthoff,  Alex Maloney, Juan Maldacena,  Don Marolf, Brian McPeak,  Ben Safdi, Douglas Stanford, and Edward Witten for valuable discussions. We are especially thankful to Jeevan Chandra for detailed comments. We thank the authors of \cite{Maldacena:2026ImaginaryDistanceBound} for sharing a draft of their paper with us. LVI and GDU were supported in part by the Leinweber Institute for Theoretical Physics
at UC Berkeley, by the Department of Energy, Office of Science, Office of High Energy
Physics through the award DE-SC0025522, and by the Department of
Energy through QuantISED award DE-SC0019380 .LVI was supported by the DOE Early Career Award DE-SC0025522. GDU is supported  by the Japan Science and Technology Agency (JST) as part of Adopting Sustainable Partnerships for Innovative Research Ecosystem (ASPIRE), Grant No.\ JPMJAP2318. CY was supported by the Simons Investigator in Physics Award MP-SIP-0001737 and U.S. Department of Energy grant DE-SC0007870. We would also like to thank Google DeepMind for offering access to the DeepThink function, and OpenAI and Anthropic for offering Pro and Max subscriptions.
\end{acknowledgments}

\appendix

\onecolumngrid

\section{Positivity constraints for the gravitational path integral}
\label{positivity}

In this appendix, we discuss the positivity constraints for the gravitational path integral and show they translate into properties of a putative statistical ensemble of Hamiltonians.  
To start, consider the inner product between two open universe states with boundary conditions $\mathcal{J}_i$ which include a Euclidean time evolution $\beta_i$, meaning the open universe boundary has topology $\Sigma_i=I_{\beta_i}\times M_{d-1}$. The inner product between two such states must be positive definite as a kernel in the boundary conditions:
\bea
 \langle \cJ_1,\beta_1|\cJ_2,\beta_2\rangle = Z_{\mathcal{J}}(\beta_1+\beta_2)\succeq0
\eea
where we denoted the result of the GPI as $Z_{\mathcal J}(\beta_1+\beta_2)$ where $\mathcal{J}=\cJ_1^\Theta \cup \cJ_2$ and the  two intervals are glued as $I_{\beta_1}\cup I_{\beta_2}=S^1_{\beta_1+\beta_2}$. It can  be proven that the following are exactly equivalent positivity conditions: 
\begin{enumerate}[label=(\arabic*)]
\item $Z_{\mathcal J}(\beta_1+\beta_2)\succeq0$; 
\item $\overline{\rho_{\mathcal J}(E)}\geq0$ where $Z_{\mathcal J}(\beta)=\int \mathrm{d}E\,\overline{\rho_{\mathcal J}(E)}e^{-\beta E}$.
\item $Z_{\mathcal J}(\beta)$ is a completely monotonic function: $(-1)^n\partial^nZ_{\mathcal J}(\beta)\geq0$ for all non-negative integer $n$. 
\end{enumerate}
We have that 
(2)$\Leftrightarrow$(3) thanks to the Bernstein–Widder theorem. We can prove that  (1)$\Leftrightarrow$(2) because $e^{-\beta E}>0$ and $\rho(E)\geq0$ if and only if the kernel is positive:
\bea
\int \mathrm{d}\beta_1\, \mathrm{d}\beta_2\, \phi^*(\beta_1) \phi(\beta_2)Z_{\mathcal J}(\beta_1+\beta_2) =\int \mathrm{d}E\,\mathrm{d}\beta_1\,\mathrm{d}\beta_2\,\overline{\rho_{\mathcal J}(E)}\,e^{-(\beta_1+\beta_2)E}\phi^*(\beta_1) \phi(\beta_2)\geq0\quad\quad\forall\phi.
\eea
More generally if we have several disconnected pieces in the open universe, each with topology $I\times S^{d-1}$, positivity states:
\bea
Z_{\mathcal{J}}(\beta_1+\tilde{\beta}_1,\ldots,\beta_n+\tilde{\beta}_n)\succeq0.
\eea
Then the following three statements are equivalent to each other: 
\begin{enumerate}[label=(\arabic*)]
\item $Z_{\mathcal{J}}(\beta_1+\tilde{\beta}_1,\ldots,\beta_n+\tilde{\beta}_n)\succeq0$
\item $\overline{\rho_{\mathcal{J}_1}(E_1) \dots \rho_{\mathcal{J}_n}(E_n)}\geq0$
\item $Z_{\mathcal{J}}(\beta_1,\dots,\beta_n)$ is a completely monotonic with respect to each argument. 
\end{enumerate}
If the Hamiltonian is not Hermitian, for example, after analytic continuation of boundary conditions $\mathcal{J}$, the same condition (2) can be formulated in terms of a generalized density of complex energy levels
\bea
\overline{\rho(E_R,E_I)}\geq 0,\qquad \tilde{Z}_{\mathcal{J}}(\beta_R,\beta_I)=\int \mathrm{d}E_R\, \mathrm{d}E_I\,\overline{\rho(E_R,E_I)}e^{-\beta_RE_R-i\beta_IE_I}.
\eea
When PT symmetry is unbroken and the support lies only on real energies this reduces to the familiar condition. Condition (2) for arbitrary number of universes implies that we have a probabilistic ensemble of theories with density $\rho_\mathcal{J}^{\alpha}$ with mean density $\overline{\rho_{\mathcal J}(E)}\geq0$ and moments $\overline{\rho_{\mathcal{J}_1}(E_1) \dots \rho_{\mathcal{J}_n}(E_n)}$, with positive probability measure:
\es{}{
\overline{\rho_{\mathcal{J}_1}(E_1) \dots \rho_{\mathcal{J}_n}(E_n)}=\int \mathrm{d}\alpha\, P(\alpha) \rho^{\alpha}_{\mathcal J}(E_1)\dots \rho^{\alpha}_{\mathcal J}(E_n), \qquad  P(\alpha)\geq 0.
}
The partition function of a single theory $Z_{\mathcal{J}}^{\alpha}(\beta)=\int \mathrm{d}E\, e^{-\beta E}\rho^{\alpha}_{\mathcal J}(E)$ is then a positive random variable $Z_{\mathcal{J}}^{\alpha}(\beta)>0$ whose moments then must satisfy the Stieltjes moment problem as described in the main text. 

\section{Details of axion wormholes}
\label{solution}
In this appendix, we discuss how we get the fixed-flux wormhole action. Starting from a wormhole with metric 
\bea
\mathrm{d}s^2=\mathrm{d}r^2 +b(r)^2 \mathrm{d}\Sigma^2_{d}
\eea
and action \eqref{eq:action}. The corresponding FRW equation is given by 
\bea
b'(r)^2=k +\frac{b^2}{\ell^2}- \frac{\tilde{C}^2}{b^{2d-2}}=F(b), \quad\quad \tilde{C}^2=C^2\frac{8\pi G_N}{d(d-1)}\label{frw}
\eea
where $k=1$ for spherical slicing, $k=0$ for torus slicing, and $k=-1$ for hyperbolic slicing. In order for the wormhole to exist, the throat size needs to have a minimum, which we denote by $b_{min}$; such a minimum exists for any slicing in AdS and for spherical slicing in flat space.  Such solutions only exist if $\tilde C >0$ (or $C>0$), in which case the axion has to be purely imaginary, or complex with a constant real part. Generally, the imaginary difference of the axion on the two asymptotic boundaries is restricted to
\bea
\Im\Delta a= |C|\Delta\tau=2|C|\int_{b_{\min}}^{\infty}  \frac{\mathrm{d}b}{b^d \sqrt{F(b)}} =
\frac{\pi }{f_a \sqrt{8\pi G_N}} \times \begin{cases}
    [0,\sqrt{\frac{d-1}{d}}) \qquad &(k=-1 \text{ in AdS}_{d+1})\\
    \\
   \sqrt{\frac{d-1}{{}d}} \qquad &(k=0 \text{ in AdS}_{d+1})\\
    \\
    \left(  \sqrt{\frac{d-1}{{}d}}, \sqrt{\frac{d}{{}d-1}} \right) \qquad &(k=1 \text{ in AdS}_{d+1}) \\
      \sqrt{\frac{d}{{}d-1}} \qquad &(k=1 \text{ in flat space}_{d+1})\\
\end{cases}
\label{eq:axion-difference}
\eea
where we note that the axion difference is fixed for $k=0$ in AdS$_{d+1}$ and $k=1$ in flat space. 
We can also simplify the Euclidean gravitational action using the FRW equation:
\begin{align}
S_\textrm{EH}+S_\textrm{GHY}&=-\frac{1}{16\pi G}\int \mathrm{d}^{d+1}x\,\sqrt{g}(R-2\Lambda)-\frac{1}{8\pi G}\int_\partial \mathrm{d}^d x \sqrt{h} K\\
&=\underbrace{-\frac{V_\Sigma}{16\pi G}\int_{-\infty}^\infty \mathrm{d}r\left(2d(d-1)kb^{d-2}-4\Lambda b^d\right)}_{S^\textrm{div}_\textrm{grav}}\underbrace{+\frac{V_\Sigma C^2}{2}\int_{-\infty}^\infty \frac{\mathrm{d}r}{b^d}}_{S^\textrm{ren}_\textrm{grav}}\label{EH+GHY}
\end{align}
The divergence is removed by background subtraction in the flat space case and by a counterterm in AdS case. Then the remaining piece is given by
\bea
S_\textrm{grav}^\textrm{ren}=\frac{V_\Sigma}{2}C^2\Delta \tau\label{gravaction}
\eea
Below, we analyze explicitly the cases where $\Delta a$  is fixed in \eqref{eq:axion-difference}, making the on-shell action in \eqref{EH+GHY} easily analytically tractable. To obtain the full action, we also have to evaluate the matter term.  This term can alternatively be described in terms of a $(D-2)$-form gauge field $B_{D-2}$ \cite{Witten:2026twr}, and we denote as $H_{D-1}=\mathrm{d}B_{D-2}$ the corresponding $(D-1)$-form field strength. The matter action is then given by
\bea
S_H=\int \mathrm{d}^D x\,\sqrt{g}\,\frac{1}{8\pi^2 d!f_a^2}H^2
\eea
$H$ satisfies
\bea
\frac{1}{2\pi}\int_{\Sigma_d}H\equiv m\in \Z
\eea
so we have
\bea
H=\frac{2\pi m}{V_\Sigma b^d}\text{vol}_\Sigma\quad\quad H^2=d!\left(\frac{2\pi m}{V_{\Sigma}b^d}\right)^2\label{Handflux}
\eea
where $\text{vol}_\Sigma$ is the volume form. Thus we have
\bea
S_H=\frac{m^2}{2f_a^2V_\Sigma}\Delta\tau
\eea
It turns out that $m$ is the conserved charge corresponding to shift symmetry of $a$ and $C=-\frac{m}{f_aV_{\Sigma}}$. Adding up all the terms in the action, we find that all terms that are non-linear in the flux $m$ cancel, leaving  
\bea
    S_\textrm{tot}(m)=S^\textrm{ren}_\textrm{grav}+S_H=V_{\Sigma}C^2\Delta\tau=\begin{cases}&\frac{|m|}{f_a} \sqrt{\frac{\pi d}{8(d-1)\,G_N}}\quad\text{flat}\\&\frac{|m|}{f_a} \sqrt{\frac{\pi (d-1)}{8d\,G_N}}\quad\quad\text{AdS}\end{cases}=|m|(\Im\Delta a_{crit})
\eea
 The fact that the action is linear in $m$ in such cases could have been deduced from the fact that the axion difference is fixed and independent of the flux in \eqref{eq:axion-difference}. This is because the boundary term necessary to go between fixed flux and Dirichlet boundary conditions for the axion is also linear in $m$ and the variation of the action with respect to $m$ determines the saddle point value of $\Delta a$. Furthermore, because $k$ can be neglected in \eqref{frw} when $C$ is large, the value of the action is independent of different slicings at leading order in the large $|m|$ expansion.

\section{Axio-dilaton wormholes in Type IIB SUGRA}
\label{axiodilaton}

We now turn to axio-dilaton wormholes.  A convenient bottom-up Euclidean action
for the axion $a$ and dilaton $\varphi$ is
\bea 
 S=\frac{1}{16\pi G_N} \int \mathrm{d}^{D}x \sqrt{g}\left(-(R-2\Lambda)%
 +\frac{1}{2}\left((\partial_\mu \varphi)^2- e^{-\beta\vp}(\partial_\mu a)^2\right)\right)
 \label{eq:axion-dilaton-action}
\eea
where we work in the convention where $f_a =\frac{1}{\sqrt{16\pi G_N}}$. 
As in Sec.~\ref{sec:wormhole-amplitudes}, there are no axion wormhole saddles with Dirichlet boundary conditions and real values for the axion and dilaton. The analogous calculation to that in Sec.~\ref{sec:wormhole-amplitudes} is to consider a fixed flux for the $(d-1)$-form field dual to the axion and Dirichlet boundary conditions for the dilaton.  Once again, we can obtain the saddles at fixed flux and their on-shell action by continuing to the wrong-sign axion variable $A=i a$. In terms of $A$, the scalar
moduli-space metric is locally
\bea
\mathrm{d} s^2_{\rm mod}
=
\mathrm{d} \varphi^2-e^{-\beta\varphi}\mathrm{d}A^2\,.
\label{eq:moduli-space-metric}
\eea
This is AdS$_2$ in Poincar\'e coordinates, with radius
$2/\beta$. %
For the low-energy limit of type IIB string theory, $\beta_\text{IIB}=-2$.

The profile of the axion and dilaton along a wormhole follows timelike geodesics in this
Lorentzian moduli space.  The scale factor of the wormhole satisfies the same equation as in \eqref{frw} with $C$ determined by the conserved energy along the moduli space geodesic,
\bea
\dot{\varphi}^2-q^2e^{\beta\varphi}=-16\pi G_NC^2,
\label{eq:conservation-of-energy}
\eea
where $q$ represents the flux of the axion along the wormhole. The distance that the geodesic travels is determined as in \eqref{eq:axion-difference}. As before, for AdS wormholes with flat slices or for flat space wormholes with spherical slices, the geodesic length in moduli space is fixed and is given by
\bea
\mathcal D_{\rm AdS}
=
\pi\sqrt{\frac{2(d-1)}{d}}\,, \qquad \mathcal D_{\rm flat}
=
\pi\sqrt{\frac{2d}{d-1}}.
\label{eq:geodesic-distance}
\eea

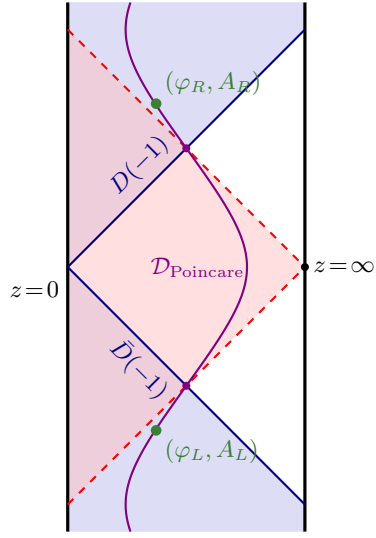
\begin{figure}
\centering
\begin{tikzpicture}[scale=1]
\def\taumax{2.5}
\def\sigmax{3.14159}
\def\smax{3.14159}
\def\tmax{3.14159}
\def\vgeo{0.72}
\def\degtorad{0.0174533}

\def\Tmax{3.5}

\fill[blue!6] (0,-\Tmax) rectangle (\smax, \Tmax);

\fill[blue!50!black, opacity=0.08] (0,0) -- (0,\Tmax) -- (\smax,\Tmax) -- ({\smax},{\smax}) -- cycle;
\fill[blue!50!black, opacity=0.08] (0,0) -- (0,-\Tmax) -- (\smax,-\Tmax) -- ({\smax},{-\smax}) -- cycle;
\fill[white] (0,0) -- ({\smax}, {\smax}) -- ({\smax}, {-\smax}) -- cycle;
\fill[pink, opacity=0.5] (0, -\smax) -- (0, \smax) -- (\smax, 0) -- cycle;

\draw[very thick] (0, -\Tmax) -- (0, \Tmax);%
\draw[very thick] (\smax, -\Tmax) -- (\smax, \Tmax);%

\draw[thick, red, dashed] (0, \smax) -- (\smax, 0);%
\draw[thick, red, dashed] (0, -\smax) -- (\smax, 0);%

\draw[thick, blue!50!black] (0,0) -- ({\smax}, {\smax})
  node[midway, above left, rotate=45] {$D(-1)$};
\draw[thick, blue!50!black] (0,0) -- ({\smax}, {-\smax})
  node[midway, below left, rotate=-45] {$\bar{D}(-1)$};

\draw[thick, violet]
  plot[domain=-\Tmax:\Tmax, samples=160, variable=\T]
  ({\smax/2+\degtorad*asin(\vgeo*cos(\T r))}, {\T});
\fill[violet] ({\smax/2}, {\smax/2}) circle (1.5pt);
\fill[violet] ({\smax/2}, {-\smax/2}) circle (1.5pt);
\fill[OliveGreen] ({\smax/2-.4}, {\smax/2+.59}) circle (2.0pt)
node[above right] {$(\varphi_R, A_R)$};
\fill[OliveGreen] ({\smax/2-.4}, {-\smax/2-.59}) circle (2.0pt)
node[below right] {$(\varphi_L, A_L)$};
\node[violet, font=\small] at ({\smax/2+0.15}, 0) {$\mathcal{D}_\text{Poincare}$};

\fill (\smax, 0) circle (1.5pt) node[right] {$z\!=\!\infty$};

\node[left, font=\small] at (0, -0.3) {$z\!=\!0$};

\end{tikzpicture}
\caption{The Lorentzian Poincare patch with metric $\mathrm{d}s^2_{\rm mod}=\mathrm{d}\varphi^2-e^{-\beta\varphi}\mathrm{d}A^2=\frac{4}{\beta^2}\frac{\mathrm{d}z^2-\mathrm{d}t^2}{z^2}$ where $t=\frac{\beta}{2}A$ and $z=e^{\beta\varphi/2}$. We have indicated where the $D(-1)$ instanton and the $\bar{D}(-1)$ anti-instanton condense; this coincides with the Bogomolny bound on the analytic continuation of $\theta$. Here $z\to \infty$ is strong coupling, whereas $z\to 0$ is weak coupling. The violet curve is a timelike geodesic in global AdS$_2$, that exits the original Poincar\'e patch. The dark green dots represent the start and end points for the moduli space trajectory in the axio-dilaton wormhole in Type IIB string theory.}
\end{figure}

An important question is whether the axion-dilaton profile described leads to a valid wormhole solution. This is because the moduli space metric \eqref{eq:moduli-space-metric} is that of the Poincar\'e patch, where timelike geodesics can only have a maximum length, 
\bea
\mathcal D_{\rm Poincar\acute{e}}
=
\frac{2\pi}{\beta}\,.
\label{eq:Poincare-patch-distance}
\eea
Traditionally, if the distance in \eqref{eq:geodesic-distance} satisfies $\cD > \cD_{\rm Poincar\acute{e}}$ then the wormhole saddle was believed to be invalid \cite{ArkaniHamed:2007js}; in such a case, the dilaton $\varphi \to -\infty$ at the Poincar\'e horizon. However, trajectories in moduli space that exceed the maximum distance in a single Poincar\'e patch are still valid solutions in the universal cover of AdS$_2$. Even if $\varphi $ diverges, the conservation of energy along the geodesic yields a finite on-shell action. To see this, we will again explicitly evaluate the action. The gravitational part of the action is the same as in \eqref{EH+GHY}. The matter action is given by
\begin{align}
S_\text{matter}&=\frac{1}{32\pi G_N}\int \mathrm{d}^{d+1}x\,\sqrt{g}(\partial\varphi)^2+\int \mathrm{d}^{d+1}x\,\sqrt{g}\frac{1}{8\pi^2 d!f_a^2}e^{\beta\varphi}H^2\\
&=\frac{V_\Sigma}{16\pi G_N}\int \mathrm{d}\tau\,\left(\frac{1}{2}\dot{\varphi}^2+\frac{1}{2}q^2e^{\beta\varphi}\right)
\end{align}
From \eqref{eq:conservation-of-energy}, we have
\bea
S_{\varphi}-S_H=-\frac{V_\Sigma}{2}C^2\Delta\tau
\eea
Adding the gravitational action from \eqref{gravaction}, we find the total on-shell action %
\bea
S_\text{axio-dilaton}=2S_H=\frac{V_\Sigma}{16\pi G_N} \int \mathrm{d}\tau\,q^2e^{\beta\varphi}=|m|\left|\Delta A_\text{crit}\right|
\label{eq:axio-dilaton-wormhole-on-shell}
\eea
where 
\bea
 |\Delta A_\text{crit}|=\frac{2}{\beta}\sqrt{2e^{\beta\vp_+}\qty[\cosh(\frac{\beta \Delta\vp}{2})-\cos(\frac{\beta}{2}\cD)]},
\eea
where $\varphi_+ = \varphi_L + \varphi_R$,  $\Delta \varphi =  \varphi_L - \varphi_R$ and $\varphi_{L,R}$ are the boundary values of the dilaton on the left and right boundaries of the wormhole.
Thus, if $\mathcal D>\mathcal D_{\rm Poincar\acute{e}}$, the wormhole
trajectory exits the original Poincar\'e patch of the axio-dilaton moduli
space, yet the resulting wormhole still yields a seemingly valid contribution. As in App.~\ref{solution}, Eq.~\eqref{eq:axio-dilaton-wormhole-on-shell} determines the leading behavior of the on-shell action at large $m$ for any wormhole slicing. This behavior determines the off-shell wormhole amplitude with Dirichlet boundary conditions, 
\bea 
Z_{WH}(\Delta A) = \sum_{m \in \mZ} e^{-|m||\Delta A_\text{crit}|} e^{i m \Delta a}\,,
\eea
where we again observe a divergence as $|\Im \,\Delta a| \to |\Delta A_\text{crit}| $. As before, this divergence, together with the fact that the one-boundary saddle is still invariant under shifts of the boundary value of the axion, leads to a violation of positivity. Thus, a theory of supergravity like \eqref{eq:axion-dilaton-action} does not lead to a well-defined Hilbert space. As long as the wormhole is perturbatively stable (see \cite{Hertog:2024nys, Marolf:2025StabilityRevisited} for an analysis in flat space), the negativity can only be resolved if the wormhole is non-perturbatively unstable. 

In Type IIB string theory the non-perturbative effects responsible for breaking the shift symmetry of the axion come from D$(-1)$-branes. More precisely, these break the $SL(2,\mathbb{R})$ symmetry of the leading order Lagrangian to $SL(2,\mathbb{Z})$.
Depending on whether we include D$(-1)$ instantons or anti-instantons, the action is 
\bea 
S_{D(-1)}^{\pm} = 2\pi\, e^{-\varphi} \pm 2\pi A\,.
\eea
The moduli space locus where  $S_{D(-1)}^{\pm} = 0$ and where the contribution of instantons is no longer suppressed is precisely at the Poincar\'e horizon of the Poincar\'e patch that describes the moduli space.
Requiring this D$(-1)$ instability to occur before the wormhole amplitude diverges thus requires us to restrict to wormholes whose trajectory in moduli space cannot be fully contained within a single Poincar\'e patch. Using \eqref{eq:geodesic-distance} and \eqref{eq:Poincare-patch-distance}, this gives a bound on $\beta$, 
\bea
\beta>\beta_{\rm crit}\,,
\qquad
\beta_{\rm crit}
=
\frac{2\pi}{\mathcal D_{\rm AdS}}
=
\sqrt{\frac{2d}{d-1}}\,.
\label{eq:beta-critical-bound}
\eea
We can easily check that this inequality is satisfied for type IIB on AdS$_5\times S^5$ ($d=4$):
\bea
\beta_{\rm crit}=\sqrt{\frac{8}{3}} < \beta_\text{IIB} = 2  %
\eea
Thus, the type IIB value of $\beta$ satisfies the requirement that follows from positivity: the
D$(-1)$-brane instability appears before the axio-dilaton wormhole can produce
an unbounded two-boundary amplitude. Note that the imaginary value of the axion at which the D$(-1)$-instantons become important precisely agrees with the Bogomolny bound in $\mathcal N=4$ SYM, the theory dual to type IIB on AdS$_5\times S^5$. At the Bogomolny bound, several one-boundary observables, such as the integrated correlators studied in \cite{Dorigoni:2021guq, Collier:2022emf}, diverge when the Bogomolny bound is saturated. Thus, the proliferation of D$(-1)$-instantons when the Bogomolny bound is saturated leads to a divergence of the one-boundary answer which would be interesting to replicate from a bulk calculation that includes the backreaction of the instantons.

\twocolumngrid 

\bibliographystyle{apsrev}
\bibliography{Biblio}

\end{document}